\documentclass[12pt]{iopart}
\usepackage{epsfig,graphics,amssymb,url,subeqnarray,color}

\begin{document}
\def\Re{{\rm Re}}
\def\v{\vspace{1cm}}
\title[Optimal  propulsive flapping in Stokes flows]
{Optimal  propulsive flapping in Stokes flows}

\author{Lo\"ic Was$^1$ and Eric Lauga$^{2}$}

\address{$^1$Department of Mechanical and Aerospace Engineering, 
University of California San Diego, 9500 Gilman Drive, La Jolla CA 92093-0411, USA. $^2$Department of Applied Mathematics and Theoretical Physics, Centre for Mathematical Sciences,  University of Cambridge, Wilberforce Road, Cambridge
CB3 0WA, United Kingdom.}

\ead{e.lauga@damtp.cam.ac.uk}
\begin{abstract}

Swimming  fish and flying insects   use the flapping of fins and wings to generate thrust. In contrast, microscopic organisms typically deform their appendages in a wavelike fashion. Since a flapping motion with two degrees of freedom is able, in theory, to produce net forces from a time-periodic actuation at all Reynolds number, we compute in this paper the  optimal flapping kinematics of a rigid spheroid in a Stokes flow. The hydrodynamics for the force generation and energetics of  the flapping motion is solved exactly. We then  compute analytically the gradient of a flapping efficiency in the space of all flapping gaits and employ it to derive numerically the optimal flapping kinematics as a function of the shape of the flapper and the amplitude of the motion. The kinematics of optimal flapping are observed to depend  weakly on the flapper shape and are very similar to the figure-eight motion observed in the motion of insect wings. Our results suggest that flapping could be a exploited experimentally as a  propulsion mechanism valid across the whole range of Reynolds numbers.

\end{abstract}

\maketitle

\section{Introduction}

Low-Reynolds number (Re) locomotion is a area of fluid mechanics where opportunities arise to pose optimization problems primarily with two motivations \cite{lauga2009}.  Deriving the shape and swimming gaits of optimal swimmers and comparing them with  experimental observations enables a direct probe of the energetic and mechanical constraints in cellular locomotion and transport \cite{spa_prl,eloylauga12}. 
In addition, as synthetic micro-swimming devices are developed for therapeutic and diagnostic tasks, the potential use of optimal swimmers would enable both cost-effective design and high-efficiency performance \cite{abbot,nelson}.

Theoretical work on optimal locomotion started with investigations on singly flagellated eukaryotic cells. The swimming gait minimizing the energy dissipation  in the fluid -- usually framed in terms of a hydrodynamic efficiency, see Refs.~\cite{lighthill75,chattopadhyay06} -- was shown to take the form of  a traveling wave \cite{pironneau74}. The shape of the hydrodynamically optimal wave was  derived by Lighthill \cite{lighthill75}, and improved upon by including energetic costs associated with flagella bending  \cite{elastic,eloylauga13}. Further work considered the parametric optimization of forms within specific, elementary, wave families \cite{pironneau74,higdon79a,higdon79b}. The optimal morphologies of model eukaryotic cells employing  one or two planar flagella were derived and successfully compared to experimental data  \cite{tamPhD,tam2011}. Similar work addressed the optimal shapes of  helical flagella, as relevant to the dynamics of bacteria \cite{fujita01,spa_prl}, and of wall-anchored flexible filaments as relevant to the dynamics of cilia \cite{Osterman2011,eloylauga12}.

In the realm of synthetic locomotion, a lot of  work  has addressed the optimal swimming problem from a theoretical standpoint. Purcell's three-link swimmer \cite{purcell77} was optimized    \cite{becker03,tam07,avron08}, and so were a simpler version of the swimmer using three aligned spheres  \cite{najafi04,alouges08}. Swimming in two dimensions, which is amenable to a  formulation using complex variables, was formally optimized  \cite{avron04:opt}. Numerical computations were used to optimize the  
deformation of synthetic cilia   \cite{tabata02}. The problem of hydrodynamically optimal locomotion and feeding by surface distortion in three dimensions was recently addressed using the squirmer model \cite{michelin_swim,michelin_feed}.

As new small-scale synthetic swimmer designs become experimental realities \cite{dreyfus05_nature,tierno08,tierno10,sing,zhang09b,ghosh,zhang10,gao,wei} there is both fundamental and practical interest in exploring the parameter space of the simplest designs possible. Because of Purcell's scallop theorem, any synthetic swimming device, or more generally, any method used to generate propulsive forces and do work at low Re under a time-periodic forcing  \cite{darnton04,kim04_diffusion,weibel05,raz08,sokolovPNAS}, needs to posses at least two degrees of freedom  and  to  actuate them in a non-time-reversible manner \cite{purcell77}. This is exemplified by Purcell's three-link swimmer \cite{purcell77}, which exploits exactly two degrees of freedom in rotation, as well as the  three-sphere unidirectional swimmer, which uses two degrees of freedom in translation \cite{najafi04}.  A third  type of force-generating device with two degrees of freedom would have one degree of freedom in translation and one in rotation -- in other words, a solid body flapper. But as a difference with  the three-link and three-sphere swimmers ubiquitous in the Stokesian literature,  flapping is known to be an  effective method to generate propulsive forces at high Reynolds numbers, as exemplified by the wing and fin motion of flying insects swimming fish.

In this paper we thus enquire on the optimal way  to actuate a flapper to generate propulsive forces in the low Reynolds number regime. Specifically, we compute the manner in which a solid-body spheroid constrained to periodically translate along a fixed direction is able to generate the maximum time-averaged propulsive force for a fixed amount of energy dissipated in the fluid. This is equivalent to  flapping with the maximum hydrodynamic efficiency.  After posing the problem mathematically, we solve exactly for the hydrodynamics of the motion,  calculate analytically the gradient in the flapping efficiency, and use it to compute numerically the optimal flapping kinematics in the translation-rotation phase space. The resulting optimal flapping motion turns out to depend weakly on the flapper shape and  to be similar to the optimal beat kinematics of insect wings in the high-Re number regime, suggesting flapping as a robust and Re-independent force generation strategy \cite{ellington84,dudley02,wang05,berman07}.

\begin{figure}[t]
\begin{center}
\includegraphics[width=.99\textwidth]{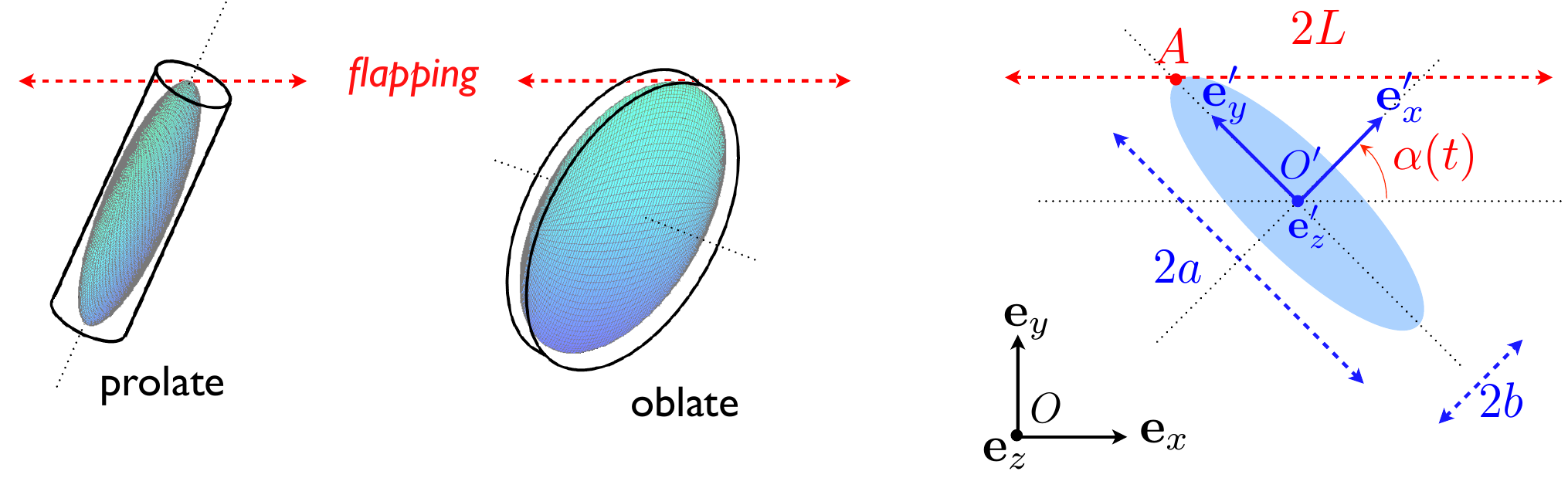}
\caption{Stokesian flapping: a rigid spheroid (either prolate or oblate) of length $2a$ and width $2b$ is periodically flapping along a prescribed straight path of length $2L$. The dynamics of the flapper angle, $\alpha(t)$, is determined as part of the optimization procedure.  Left: illustration of prolate and oblate flapping. Right:  notation (see text).}
\label{fig1}
\end{center}
\end{figure}

\section{Setup}
\subsection{Geometry}
The problem considered in the paper is illustrated in Fig.~\ref{fig1}. The extremity, A, of  solid body spheroid is flapping along a prescribed straight  path. The lab-fixed  frame is denoted $\mathcal{R}_{0}$ with center $O$ and unit vectors  $\left({\bf e}_{x},{\bf e}_{y},{\bf e}_{z}\right)$. The body-moving frame attached to the center of the spheroid is denoted  $\mathcal{R}_{1}$, with center $O'$ and unit vectors  $\left({\bf e}'_{x},{\bf e'}_{y},{\bf e'}_{z}\right)$. For a prolate spheroid, the axis of symmetry is along ${\bf e'}_{y}$, and it is along ${\bf e'}_{x}$ for an oblate spheroid, so that in both case a net force is expected to be induced in the ${\bf e}_y$ direction; in both cases
the major (resp.~minor) axis has length $2a$ (resp.~$2b$). The edge of the flapper, $A$, located at $\left(0,a,0\right)_{\mathcal{R}_{1}}$, translates  along the direction ${\bf e}_{x}$  with position $f(t)$ and rotates around  the $({\bf e}_{x},{\bf e}_{y})$ plane with angle $\alpha(t)$. The functions $f$ and $\alpha$ are time-periodic with period $\tau$. Time is chosen such that $f(0)=f(\tau)=0$, and the translation has an amplitude of $2L$, i.e.  $\max f(t) = -\min f(t) = L$. Any point ${\bf r}$ on the solid body moves thus with the instantaneous  velocity
\begin{equation}
	{\bf V}({\bf r},t)= \dot{f}(t){\bf e}_{x}+\dot{\alpha}(t){\bf e}_{z}\times{\bf r},
	\label{eq:vitesse}
\end{equation}
with dots denoting time derivatives. Due to the two degrees of motion of the flapper (translation and rotation), in general a net propulsive force will be generated along the $y$ direction (i.e.~at right angle with respect to the flapping direction). The optimal flapping problem consists in determining the  $\alpha\equiv \alpha(f)$ relationship in the translation-rotation phase space together with the rate at which the flapping gait is being performed in order  to maximize the propulsion from flapping for a fixed energetic cost (or equivalently, minimize the energetic cost of flapping for a fixed propulsion).

\subsection{Dimensionless numbers}

Two dimensionless numbers characterize the optimal flapping kinematics. The first one is the flapper aspect ratio, $\Delta$, defined as
\begin{equation}
	\Delta = \frac{\rm{Axis \,\,of\,\, symmetry\,\, length}}{\rm{Other\,\, axis\,\, length}},
	\label{eq:defdelta0}
\end{equation}
and $\Delta ={a}/{b} >1$ for prolate, $\Delta=b/a < 1$ for oblate, 1 for a sphere. The limit  $\Delta\rightarrow+\infty$ corresponds to an elongated rod, while for $\Delta\rightarrow0$ it is a flattened disk. The other dimensionless number is the  flapping amplitude, $\Pi$, which we define as $\Pi = {L}/{a}$.

\subsection{Dynamics}
Let us denote by  ${\bf F}$ and ${\bf T}$ the net force and the torque applied by the fluid on the moving body; torques are evaluated at the center $O'$ of the spheroid. If $\eta$ denotes the shear viscosity of the fluid, then in the Stokes flow regime ($\Re=0$) we have
\begin{equation}
	\left(
	\begin{array}{c}
		{\bf F}\\
		{\bf T}
	\end{array}\right)
	=-\eta
	\left(
	\begin{array}{cc}
		{\bf {\bf A}} & {\bf {\bf B}}\\
		^{t}{\bf {\bf B}} & {\bf {\bf C}}
	\end{array}\right)\cdot
	\left(
	\begin{array}{c}
		{\bf U}\\
		{\bf \Omega}
	\end{array}\right),
\label{eq:relefforts}
\end{equation}
where ${\bf {\bf A}}$ and ${\bf {\bf C}}$ are symmetric resistance tensors \cite{bi:Guyon,happel}, ${\bf \Omega}$ the rotation rate of the solid body, and ${\bf U}$ the translation velocity of the spheroid center. Since spheroids have three orthogonal planes of symmetry  we have ${\bf {\bf B}}={\bf 0}$, and therefore
$	{\bf F}=-\eta{\bf {\bf A}}\cdot{\bf U} $ and $ {\bf T}=-\eta{\bf {\bf C}}\cdot{\bf \Omega}
$. In the body frame $\mathcal{R}_{1}$, ${\bf {\bf A}}$ and ${\bf {\bf C}}$ are diagonal \cite{bi:Guyon,happel},  and we write
 ${\bf A} 
 = \lambda_{\perp} {\bf e}_x' {\bf e}_x' 
 + \lambda_{\parallel} {\bf e}_y' {\bf e}_y' 
 + \lambda_{z'} {\bf e}_z' {\bf e}'_z$
and 
 ${\bf C} 
 = \gamma_{\perp} {\bf e}_x' {\bf e}_x' 
 + \gamma_{\parallel} {\bf e}_y' {\bf e}_y' 
 + \gamma_{z'} {\bf e}_z' {\bf e}_z'$
where $\left[\lambda_{z'},\gamma_{z'}\right]=\left[\lambda_{\perp},\gamma_{\perp}\right]$ for a {prolate} spheroid and $\left[\lambda_{z'},\gamma_{z'}\right]=\left[\lambda_{\parallel},\gamma_{\parallel}\right]$ for an {oblate} spheroid. The values for the  individual resistance coefficients, $\lambda$'s, are known exactly \cite{kimbook} and we always have 
$\lambda_{\perp} \geq \lambda_{\parallel}$.

Given Eq.~(\ref{eq:vitesse}), it is straightforward to evaluate the  force,  ${\bf F}_{f}$, and torque,  ${\bf T}_{f}$, applied by the flapping motion on the surrounding fluid, and we obtain in the laboratory frame
\begin{equation}
	{\bf F}_{f}(t)=\left(
	\begin{array}{c}
		\eta\dot{f}(t)\left[\lambda_{\perp}\cos^{2}\alpha(t)+\lambda_{\parallel}\sin^{2} \alpha(t)\right]\\
		\eta\left(\lambda_{\perp}-\lambda_{\parallel}\right)\dot{f}(t)\cos \alpha(t) \sin \alpha(t)\\
		0
	\end{array}\right)_{\mathcal{R}_{0}}
\label{netforce}\end{equation}
and
${\bf T}_{f}(t)= \eta\gamma_{z'}\dot{\alpha}(t){\bf e}_z$.
We notice that drag anisotropy ($\lambda_{\perp}\neq\lambda_{\parallel}$), which occurs for any non-spherical spheroid, leads to a nonzero propulsive force being generated in the direction  perpendicular to the flapping direction ($y$).

The instantaneous rate of working, ${\dot  \mathcal{W}} \geq 0$,   of the loads applied by the flapper on the fluid is given by
\begin{equation}
	{\dot  \mathcal{W}}(t)={\bf F}_{f}(t)\cdot{\bf V}(O',t)+{\bf T}_{f}(t)\cdot{\bf \Omega}(t)
	\label{eq:formpuissance}
\end{equation}
which is
\begin{eqnarray}
{\dot \mathcal W}(t)&=&\eta\left[\dot{f}^{2}(t)\left(\lambda_{\perp}\cos^{2}\alpha(t)
+\lambda_{\parallel}\sin^{2}\alpha(t)\right)\right.\nonumber\\
&& \quad \left.+a\lambda_{\perp}\dot{f}(t)\dot{\alpha}(t)\cos \alpha(t)+\gamma_{z'}\dot{\alpha}^{2}(t)\right]\label{eq:puissance}.
\end{eqnarray}
The total work done by flapper during a period is
\begin{equation}
	W_{tot}=\int_{0}^{\tau}{\dot \mathcal W}(t)\mathrm{d}t
	\label{eq:travailtot}.
\end{equation}

\subsection{Time parametrization and minimum work}

Over a single period, the net force, $\int {\bf F}_{f}{\rm d} t$, and torque, $\int {\bf T}_{f}{\rm d} t$,  applied on the fluid is not a function of the particular rate at which  the flapping motion is taking place. This arises  because of the time-independence of Stokes equations  and can be seen by observing that both ${\bf F}_{f}(t)$ and ${\bf T}_{f}(t)$ are exact time-derivatives. As a difference, rates do affect the work done by the flapper. It was shown by Becker {\it et al.} \cite{becker03} that, for any particular time parametrization, the work done by the flapper was always above a particular minimum value given by
\begin{equation}
	W_{tot} \geq W_{min}=\frac{1}{\tau}\left[\int_{0}^{\tau}
	\sqrt{{\dot \mathcal W}(t)}\mathrm{d}t\right]^{2}
	\label{eq:travailmin},
\end{equation}
as a result of the Cauchy-Schwartz inequality. 
This statement   is actually very powerful, because 
the square root of the rate of working appearing as an integrand in the right hand side of Eq.~\ref{eq:travailmin} is an exact derivative, and thus the value of $W_{min}$ does not depend on the particular rate at which  flapping is taking place. Flapping at the minimum rate of working, $W_{tot}=W_{min}$, is obtained for the particular rate of flapping chosen such that  ${\dot \mathcal W}(t)$ remains constant during the flapping period.

\subsection{Flapping efficiency}

In this paper we want to produce maximum flapping with minimum energetic cost. Similarly to what is done in the context of low-Reynolds number swimming \cite{lauga2009,spa_prl,lighthill75,chattopadhyay06,pironneau74,higdon79a,higdon79b}, we define a flapping efficiency, $\varepsilon$, as the ratio between an effective propulsive cost, $W_{eff}$,  to the total energetic cost of flapping, 
\begin{equation}
	\varepsilon = \frac{W_{eff}}{W_{tot}},
\end{equation}
where $W_{tot}$ is defined by Eq.~\ref{eq:travailtot}. {Different definitions of $W_{eff}$ can be proposed. To render $\varepsilon$ independent of the flapping frequency, the effective propulsive work needs to scale quadratically with the propulsive force}. From a dimensional standpoint, it {should} then follow the scaling 
 $W_{eff}\sim\tau \left\langle F_{y}\right\rangle^{2} / \eta\lambda_{\parallel}$. Physically, this scaling arises from considering the typical velocity, $U$, at which a body would move in response to a net propulsive force, $\left\langle F_{y}\right\rangle$, namely 
  $U\sim \left\langle F_{y}\right\rangle/\eta \lambda$, 
 and equating the effective energy to the power $\left\langle F_{y}\right\rangle U$ times the flapping period. {We choose} $\lambda \equiv \lambda_{\parallel} $ because the  flapper is, on average,  oriented parallel to the propulsive direction, and finally get the efficiency
 \begin{equation}\label{finaleps}
	\varepsilon = \frac{\tau\left\langle F_{y}\right\rangle^{2}}{\eta\lambda_{\parallel}W_{tot}}
	\label{eq:efficacite},
\end{equation}
where the integration of Eq.~\ref{netforce} leads to
\begin{equation}
	\left\langle F_{y}\right\rangle = \frac{1}{\tau}\int_{0}^{\tau}F_{y}(t)\mathrm{d}t = \frac{\eta(\lambda_{\perp}-\lambda_{\parallel})}{2\tau}\int_{0}^{\tau}\dot{f}(t)\sin 2\alpha(t)\mathrm{d}t
	\label{eq:fymoy}.
\end{equation}
{Other choices for the value of $\lambda$ can be made but they do not change the optimal flapping  kinematics  obtained below. 
The form of $\varepsilon$ in Eq.~\ref{eq:efficacite} insures that it does not depend on the flapping frequency nor  the fluid viscosity, and therefore the efficiency} is only a function of three parameters: the flapping amplitude $\Pi$, the flapper aspect ratio $\Delta$, and the particular flapping gait. Our goal is therefore to derive the optimal flapping gait maximizing $\varepsilon$ for fixed values of both $\Pi$ and $\Delta$.


\section{Steady flapping and optimal angle}
\label{optimalangle}

A first result can be derived analytically. {Let us consider the situation where the flapping distance is large ($\Pi\gg 1$). Far from the end points on the flapping path, we expect the solid body to reach a steady configuration with a constant angle and a steady translation velocity. In other words, both $\alpha$ and $\dot f$ should be constant.} 
What is then the optimal steady flapping angle, $\alpha$, maximizing the efficiency $\varepsilon$? Intuitively, when $\alpha$ is zero or $\pi/2$ the propulsion direction ($y$) is aligned with one of the principal axis of the flapper, and therefore no propulsion occurs. An optimal angle must therefore exist. In the steady limit, the efficiency is given by
\begin{equation}
	\varepsilon = \frac{\tau\left\langle F_{y}\right\rangle^{2}}{\eta\lambda_{\parallel}W_{tot}} 
	= \frac{\eta^{2}\left(\lambda_{\perp}-\lambda_{\parallel}\right)^{2}U^{2}\cos^{2} \alpha \sin^{2} \alpha }{\eta^{2}\lambda_{\parallel} U^{2}\left(\lambda_{\perp}\cos^{2} \alpha +\lambda_{\parallel}\sin^{2} \alpha \right)} =(\beta-1)^{2}G(\alpha),
\end{equation}
where $\beta= {\lambda_{\perp}}/{\lambda_{\parallel}} $ and 
\begin{equation}
	G(\alpha) = \frac{\cos^{2} \alpha \sin^{2} \alpha }{\beta\cos^{2} \alpha +\sin^{2} \alpha }\cdot
\end{equation}
The maximum of the function $G$ is obtained for a particular angle, $\alpha_{{\rm opt}}$, given by  
\begin{equation}
\alpha_{{\rm opt}}=\arccos\left[\left(1+\sqrt{\beta}\right)^{-{1}/{2}}\right],
\end{equation}
 with a corresponding flapping efficiency of 
 \begin{equation}
\varepsilon_{max}(\beta) = \left(\sqrt{\beta}-1\right)^{2}.
\end{equation}
In the very slender limit of prolate spheroids, $a\gg b$, we have $\beta \to 2$, leading to an optimal angle $\alpha_{{\rm opt}}\approx 50^\circ$ and a flapping efficiency 
$\varepsilon_{max}\approx 17.2\%$. That angle is reminiscent of the constant flagellar slope derived by Lighthill in the context of optimal locomotion using flagellar waves \cite{lighthill75}. Everything else being equal, we see from 
Eq.~\ref{eq:fymoy} that the propulsive force is maximum when $\sin 2\alpha =1$ and thus $\alpha=\pi/4$. The actual optimal angle is larger than that  value because, from an energetic standpoint, angles above $\pi/4$ require less work (per unit speed of translation)  than those below $\pi/4$. In the opposite oblate limit of thin disks, we have $\beta \to 3/2$, leading to $\alpha_{{\rm opt}}\approx 48^\circ$ and an efficiency  $\varepsilon_{max}\approx 5\%$. 
As will be discussed below, when the flapping amplitude is large,  optimal flapping will be composed of {such} steady flapping periods near the optimal angle followed by turning events.

\section{Optimal unsteady flapping}
\subsection{Calculus of variations}

We now consider the general case of unsteady flapping. We wish to find the flapping gait leading to a maximum value of the efficiency given in Eq.~\ref{finaleps}. In order to numerically compute the optimal flapping gait, we first use calculus of variations to analytically determine the small change in the flapping efficiency, $\delta\varepsilon$,  resulting from a small change in the gait. In order to do so, let us imagine that the translational part of the flapping is a known periodic function of time, $f(t)$. We then  consider a particular time periodic orientation, $\alpha(t)$, and assume it undergoes a small change $\alpha \to \alpha + \delta \alpha$. Given Eq.~\ref{finaleps}, the resulting change in the flapping efficiency is given by 
\begin{equation}
	\delta\varepsilon  =  \frac{\tau}{\eta\lambda_{\parallel}}\frac{\left\langle F_{y}\right\rangle}{W_{tot}^{2}}\cdot\left(2W_{tot}\delta\langle F_{y}\rangle-\left\langle F_{y}\right\rangle\delta W_{tot}\right),
	\label{eq:de}
\end{equation}
and have thus to compute the change to the force, $\delta\langle F_{y}\rangle$, and the change to the work, $ \delta W_{tot}$. To calculate the first one, we use Eq.~\ref{eq:fymoy} to obtain 
\begin{equation}
	\delta\langle F_{y}\rangle = \frac{\eta(\lambda_{\perp}-\lambda_{\parallel})}{\tau}\int_{0}^{\tau}\delta\alpha\cdot\dot{f}(t)\cos[2\alpha(t)]\mathrm{d}t.
	\label{eq:dfy}
\end{equation}
The change in the total work can be found using Eqs.~\ref{eq:travailtot} and \ref{eq:puissance}, leading to
\begin{equation}
 \delta W_{tot} = \eta\int_{0}^{\tau}\left[\delta\alpha\cdot h_{1}(\dot{f},\alpha,\dot{\alpha})+\delta\dot{\alpha}\cdot H_{2}(\dot{f},\alpha,\dot{\alpha})\right]\mathrm{d}t
\end{equation}
where
\begin{eqnarray}
\displaystyle		h_{1}(\dot{f},\alpha,\dot{\alpha}) & = &\displaystyle \left(\lambda_{\parallel}-\lambda_{\perp}\right)\dot{f}^{2}\sin(2\alpha)-a\lambda_{\perp}\dot{f}\dot{\alpha}\sin(\alpha),\\
\displaystyle		H_{2}(\dot{f},\alpha,\dot{\alpha}) & = &\displaystyle a\lambda_{\perp}\dot{f}\cos(\alpha)+2\gamma_{z'}\dot{\alpha}.
\end{eqnarray}
As $f$ and $\alpha$ are both time periodic function, we can use an integration by part to obtain
\begin{equation}
 \delta W_{tot} = \eta\int_{0}^{\tau}\delta\alpha\left[ h_{1}(\dot{f},\alpha,\dot{\alpha})+ h_{2}(\dot{f},\ddot{f},\alpha,\dot{\alpha},\ddot{\alpha})\right]\mathrm{d}t,
\end{equation}
where the function $h_2$ is given by
\begin{equation}
 h_{2}(\dot{f},\ddot{f},\alpha,\dot{\alpha},\ddot{\alpha}) = \frac{\mathrm{d}H_{2}(\dot{f},\alpha,\dot{\alpha})}{\mathrm{d}t}\cdot
\end{equation}
Carrying out the algebra we obtain
\begin{eqnarray}
	\delta W_{tot} &=& \eta\int_{0}^{\tau}\delta\alpha\left[\left(\lambda_{\parallel}-\lambda_{\perp}\right)\dot{f}^{2}(t)\sin[2\alpha(t)]\right.\nonumber\\ 
	&& \quad\quad\left.-a\lambda_{\perp}\ddot{f}(t)\cos[\alpha(t)]-2\gamma_{z'}\ddot{\alpha}(t)\right]\mathrm{d}t.
	\label{eq:dW}
\end{eqnarray}
Combining Eqs.~\ref{eq:dfy} and \ref{eq:dW} into Eq.~\ref{eq:de} we finally get the gradient 
\begin{equation}
	\delta\varepsilon = \frac{\tau}{\lambda_{\parallel}}\frac{\left\langle F_{y}\right\rangle}{W_{tot}^{2}}\int_{0}^{\tau}\delta\alpha\cdot H(\dot{f},\ddot{f},\alpha,\dot{\alpha},\ddot{\alpha})\mathrm{d}t,
	\label{eq:definal}
\end{equation}
where the function $H(\dot{f},\ddot{f},\alpha,\dot{\alpha},\ddot{\alpha})$ is given by 
\begin{eqnarray}\nonumber
H &=& \frac{2W_{tot}}{\tau}(\lambda_{\perp}-\lambda_{\parallel})\dot{f}\cos 2\alpha \\
&& -\left\langle F_{y}\right\rangle\left[(\lambda_{\parallel}-\lambda_{\perp})\dot{f}^{2}\sin 2\alpha -a\lambda_{\perp}\ddot{f}\cos \alpha -2\gamma_{z'}\ddot{\alpha}\right]
,	\label{eq:fctH}\end{eqnarray}
and where $W_{tot}$ and $\left\langle F_{y}\right\rangle$ depend on the instantaneous values of $ \alpha$, $\dot{f}$, and $\dot{\alpha}$.

\subsection{Numerical implementation}

 We discretize the time interval $\left[0,\tau\right]$ in $N$ points, and both $f$ and $\alpha$ are so discretized. Their first and second derivatives with respect to time are defined using centered finite differences extended to end points using time-periodicity  of the functions.  The analytical gradient, $H$, derived in the previous section (Eq.~\ref{eq:fctH}) is then used numerically to derive the  optimal flapping motion. Two things need to be found in order to converge to the optimal solution: the optimal relationship $\alpha\equiv \alpha(f)$ and the optimal rate at which the flapping motion leading to minimum work. 

\begin{figure}[t]
\begin{center}
\includegraphics[width=.99\columnwidth]{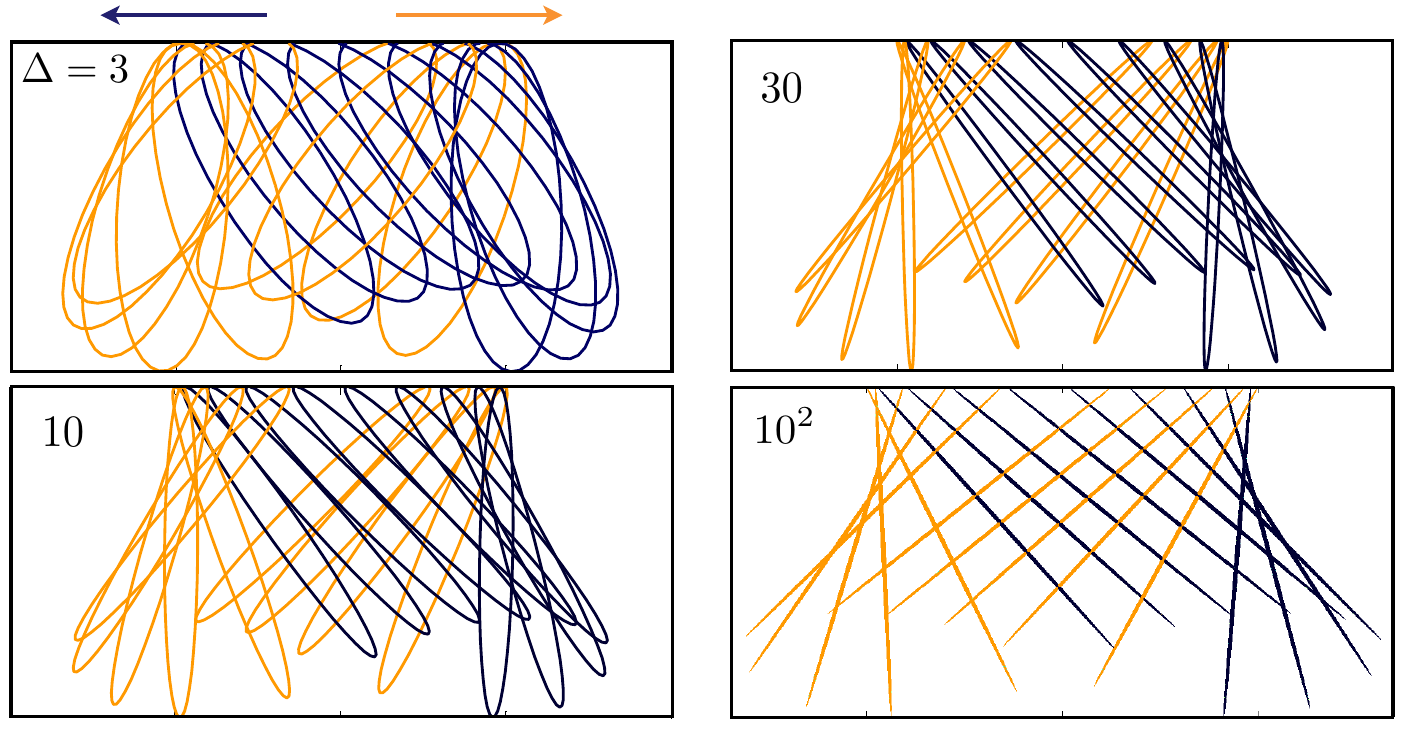}
\caption{
Flapping strokes during one period of optimal flapping with $\Pi=1$ for $\Delta=3,$ 10, 30, and $10^2$. The flapping efficiencies are respectively 0.6\%, 2.2\%, 3.6\%, and 4.8\%. The light orange shapes display strokes during the right-moving portion of the flapper hinge  while the black correspond to a left-moving hinge.}
\label{fig3}
\end{center}
\end{figure}

 We first start by a uniform discretization, termed $t_0$, of the interval $\left[0,\tau\right]$ with $N$ points. We take as initial guess the dimensionless flapping dynamics ${f_{0}} = \sin(2\pi {t_{0}})$, and numerically solve for the discrete solution, $\alpha_0$, solving  the nonlinear equation $H(\alpha)=0$ in a discretized sense using Matlab. With this  solution for both $f(t)$ and $\alpha(t)$, we find the new time-parametrization which leads to flapping with minimum rate of working, first by calculating the value of $W_{min}$ using Eq.~\ref{eq:travailmin}, and then by finding the new distribution of time intervals such that ${\dot \mathcal W}(t_{i})={W_{min}}/{\tau}$ for $i=1\dots N$. With this new time parametrization we iterate. Convergence to the optimal solution is usually obtained in less than 10 steps  {and systematically leads to a unique solution}. The value $N=100$ was found sufficient for our results to quickly converge and larger values of $N$ leave our results unchanged.

\subsection{Optimization results}

The flapping stokes resulting from the  optimization procedure are shown in  Fig.~\ref{fig3} in the prolate case for 4 slenderness ratios ($\Delta=3$, 10, 30, and $10^2$) at  
equally-spaced instants of time. The  kinematics displayed in Fig.~\ref{fig3}, {which show both translation and rotation of the flapper} and 
appear to depend very weakly on the actual flapper shape, are the main results of this paper. The free-end of the flapper describes a figure-eight motion,  a type of kinematics  surprisingly reminiscent of the  path followed by insect wings at very high Reynolds number \cite{ellington84,dudley02,wang05,berman07}. Specifically, the angle  describing the orientation seem to be ahead, from a phase standpoint, from the translation, such that when the flapper  reaches its maximum flapping amplitude and starts moving in the opposite direction, its orientation has already changed so as to produce a propulsive force in the correct direction (here, ${\bf e}_y$). 

This phase difference can be understood with  simple scaling  arguments. Examining Eq.~\ref{netforce} for the value of the propulsive force, we see that from a scaling standpoint the force induced on the fluid is 
$F_y(t) \sim   \dot f(t) \cos\alpha(t)\sin\alpha(t)$ and in the small angle limit it becomes $F_y(t)\sim  \dot f\alpha $. With a sinusoidally-varying flapping position, $f\sim \sin t$, and postulating $\alpha\sim \sin(t+\phi)$ then we see that the induced time-averaged force is given by 
$\langle F_y \rangle\sim \sin\phi$. Given that $F_y$ is the force acting from the flapper on the fluid we want it to be negative and of maximum magnitude  (in order to induce propulsion and locomotion in the positive $y$ direction), which occurs  for a flapping angle ahead   of the flapper position with
 $\phi=- \pi/2$ and thus $\alpha \sim - \cos t$.

\begin{figure}[t]
\begin{center}
\includegraphics[width=.99\columnwidth]{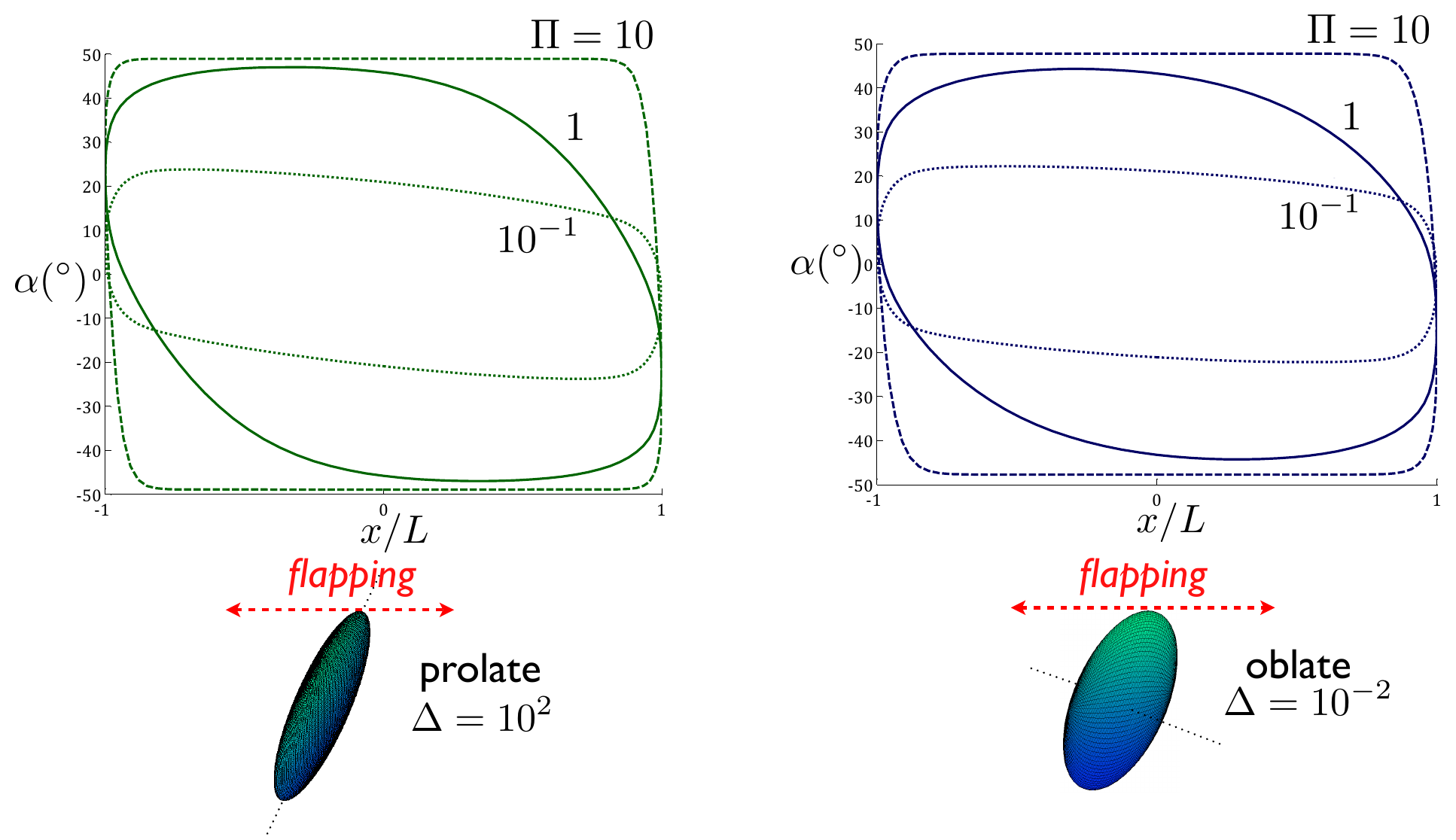}
\caption{
Optimal flapping for prolate ($ \Delta = 10^2$, left) and oblate ($ \Delta = 10^{-2}$, right) spheroids. The stroke is illustrated in the  dimensionless $\alpha$ vs.~$x/L$ space for three values of the dimensionless flapping distance $\Pi=0.1$ (dotted), 1 (solid), and 10 (dashed line).  }
\label{fig2}
\end{center}
\end{figure}

In contrast with the shape change which has very little impact on the optimal kinematics, varying the amplitude of the flapping motion  greatly influences the optimal flapping motion. This is shown in  Fig.~\ref{fig2} where we display the optimal flapping strokes in the dimensionless $\alpha$ vs.~$x/L$ plane for three flapping amplitudes ($\Pi=0.1$, 1 and 10) and both a prolate spheroid (left, $\Delta=10^2$) and an oblate one (right, $\Delta =10^{-2}$).  As can be seen by comparing the right and left sides of Fig.~\ref{fig2}, the optimal flapping path is very similar for prolate and oblate shapes, suggesting that it might be used as a robust force-generation mechanism at low Reynolds number. 
Along the paths shown in Fig.~\ref{fig2}, the kinematics occurs in the counter-clockwise direction, leading to a force induced on the fluid in the $-y$ direction, and thus a force on the flapper (and subsequent locomotion) in the $+y$ direction. {Given that the
Stokes equations are time-reversible, an equally optimal solution would rotate along the same paths but in the clockwise direction,  producing a propulsive force of identical magnitude but opposite sign. In other words, if a particular flapping kinematics is optimal with time going forward, the one obtained by going backwards in time will also be  optimal.} Note that the small-angle limit $\alpha\ll 1$ introduced above in order to understand the phase difference between flapping and pitching would predict an ellipsoidal path in parameter space.

\begin{figure}
\begin{center}
\includegraphics[width=0.54\columnwidth]{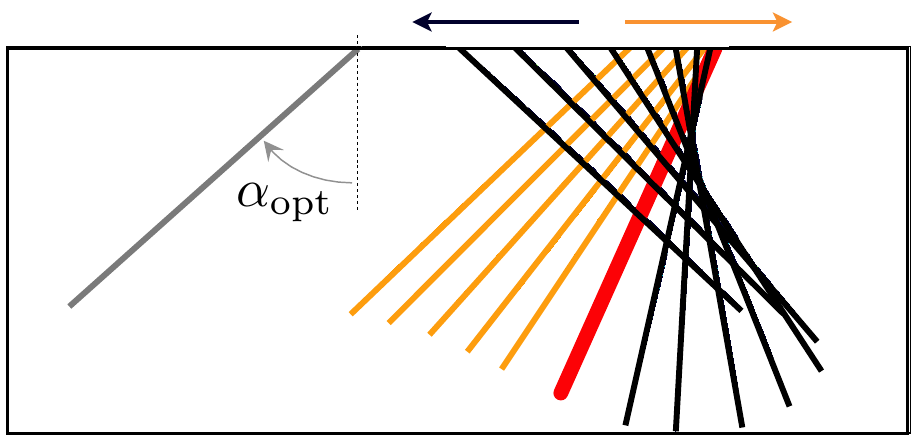}
\caption{Optimal turning  for a prolate slender flapper ($\Delta=10^2$) as the end solution of optimal flapping with $\Pi=10$. The grey shape on the left shows the optimal flapping angle in the case of steady flapping.  
During the flapping stroke, right-moving shapes are shown in light orange and left-moving shapes  in black; the red shape (thick line) corresponds to the instant where the translational velocity of the flapper is zero.}
\label{fig4}
\end{center}
\end{figure}


For large flapping amplitude ($\Pi=10$) we see that most of the path occurs at a constant flapping angle, close to the one derived in \S\ref{optimalangle}, with quick turning events. 
As a difference, when $\Pi$ is of order one or less, the flapping distance is not sufficiently long to allow the optimal angle to be reached.
A turning event in the limit of large flapping amplitude is further illustrated in Fig.~\ref{fig4}, where it is compared with the theoretical optimal flapping angle in the steady limit (grey). In addition to right- and left-moving shapes, we display in red (thick line) the particular shape at which the moving hinge of the flapper changes direction. 

\begin{figure}
\begin{center}
\includegraphics[width=0.99\columnwidth]{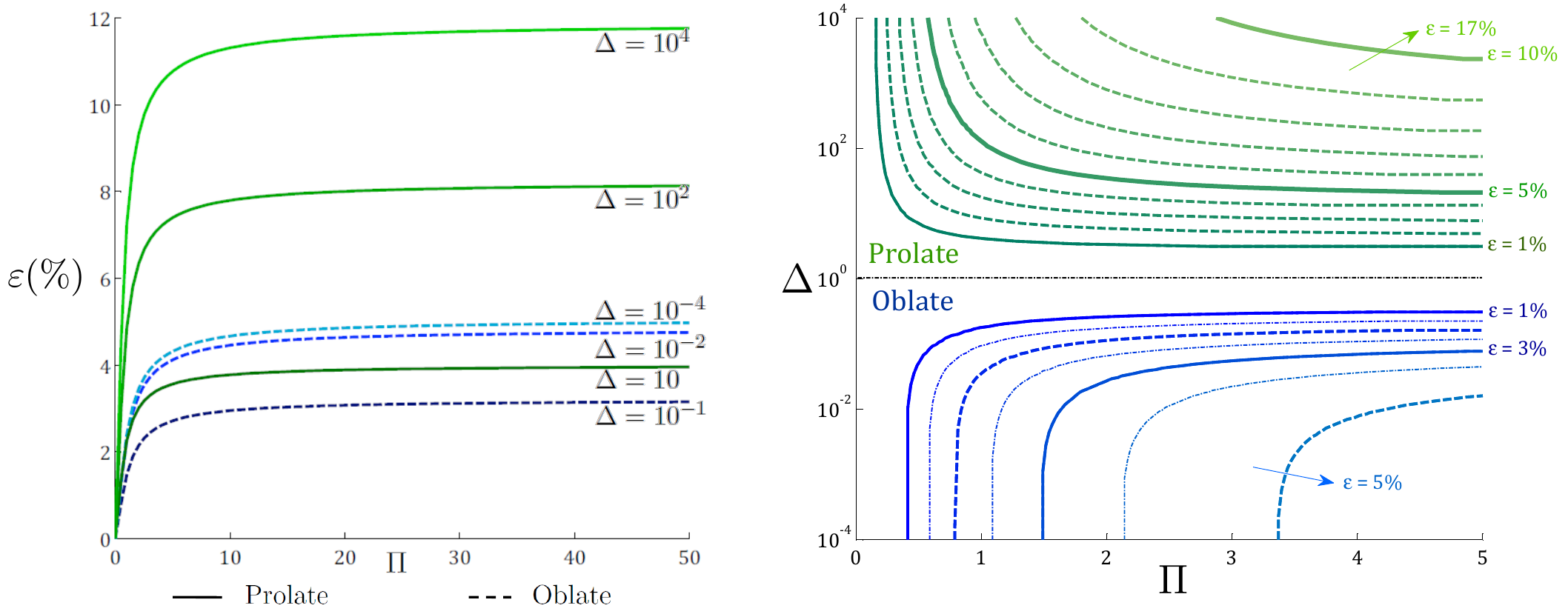}
\caption{Left: Dependence of the flapping efficiency, $\varepsilon$, on  the  flapping distance, $\Pi$, for prolate (solid lines) and oblate shape (dashed) of various aspect ratios.  Right: Iso-values of the flapping efficiency, $\varepsilon$, as a function of flapper aspect ratio ($\Delta$) and flapping distance ($\Pi$); Top:  prolate spheroid,  each line displaying increments of 1\% in efficiency; Bottom:  oblate, each line shows increments of 0.5\% in efficiency.}
\label{fig6}
\end{center}
\end{figure}

Although the difference in shape (prolate vs.~oblate) does not lead to noticeable changes in the optimal kinematics, one aspect in which both shapes differ is in their flapping efficiency. In Fig.~\ref{fig6} (left), we display the dependence of the flapping efficiency on the flapping amplitude for various prolate (solid lines) and oblate (dashed lines) shapes. For a given shape, $\Delta$, the efficiency increases monotonically with the flapping amplitude until asymptotically reaching the steady limit of \S\ref{optimalangle}. In Fig.~\ref{fig6} (right), we show the iso-values of the flapping efficiency as a function of the dimensionless flapping amplitude, $\Pi$, and the flapper shape $\Delta$, with prolate spheroids shown in the upper half (green) and oblate in the bottom half (blue). Note that spheres do not generate propulsive forces and have thus zero efficiency. Flapping using prolate shapes systematically outperforms oblate shapes of similar aspect ratio. A flapping efficiency of $\sim17\%$\footnote{{The  values of $\varepsilon$ in Fig.~\ref{fig6} should not be directly compared with the typically small efficiencies of biological microswimmers \cite{chattopadhyay06}. This is because  we do not consider free-swimming motion in our paper, but only the optimal generation of a propulsive force.}}  is the maximum possible efficiency, and it is obtained in the limit of large slenderness and flapping amplitude. In contrast, the maximum efficiency which can be reached in the limit of infinitely-thin plates is only on the order of 5\% (both are the asymptotic steady limits from \S\ref{optimalangle}).  Flappers of a fixed motion amplitude, $\Pi$, are increasingly more efficiency as they because more slender (prolate) or thin (oblate).

\section{Discussion}

In this paper we derived  the optimal flapping strokes of prolate and oblate spheroids. The focus on simple geometrical shapes allowed us to describe the hydrodynamics for the  force generation and its energetics exactly. It also allowed us  to derive analytically the gradient of the efficiency in the flapping stroke, leading to a straightforward computational implementation. The results we obtained are significant both in a biological context  and for biomimetic applications. Biologically, it is quite remarkable that the optimal kinematics for force generation at low Reynolds are so similar to the ones observed in the beat kinematic of insect wings \cite{ellington84,dudley02,wang05,berman07}, although for completely different physical reasons. The physics relevant to force generation in the flying and hovering  of insects and the swimming of fish is governed by the principles of unsteady  aerodynamics and involves vorticity dynamics in the wake and flow separation, and their relationship to the wing/fin angle of attack 
\cite{wang05}. In our case,  obviously, the physics is completely different. Instead of reactive dynamics, all the hydrodynamics forces generated at low Reynolds number  are resistive in nature, including the one inducing the net propulsion. Furthermore, the most efficient shapes in the Stokesian limit are not wings (oblate) but filaments (prolate), and the hydrodynamic efficiency of flapping {in  our low-$\Re$ study  is significantly   smaller than the typically large efficiency of high-$\Re$ flapping flight}.  Despite these differences, the typical figure-eight motion shown in the wing/filament kinematics, and the phase delay between flapping and pitching, are sufficiently  similar to be noted. Whereas a lot has been written about the fundamental differences between locomotion in the low vs.~high Re number world, the similarities noted in our paper appear to be unique. From an applied standpoint for biomimetic applications, our results suggest that flapping could be exploited as a unique Re-independent propulsion mechanism. Clearly experimental challenges exist in an  implementation of a mechanism with moving parts on  small length scales such as the one considered here. {Furthermore, the optimal flapping kinematics would have to be re-derived for any specific free-swimming implementation of the idea, taking account the force and moment distribution on the whole, free-swimming, device.} We hope that the  existence of the optimal, $\Re$-independent, motion derived in our work  will stimulate further experimental work.

\section*{Acknowledgments}
Funding by the US National Science Foundation (grant CBET-0746285 to EL) is gratefully acknowledged. 
\v
\bibliographystyle{unsrt}
\bibliography{flap_bib}
\end{document}